\documentclass[MSNbibl,nameyear,dvips]{arxstspdf}
\usepackage{flushend}
\usepackage{stfloats}

\volume{26}
\issue{2}
\pubyear{2011}
\firstpage{260}
\lastpage{261}
\doi{10.1214/11-STS346B}
\referstodoi{10.1214/10-STS346}

\begin{document}
\begin{frontmatter}

\title{Discussion of
``Impact of Frequentist and Bayesian Methods on Survey Sampling
Practice:
A Selective Appraisal'' by J.~N.~K.~Rao}
\runtitle{Discussion}
\pdftitle{Discussion of Impact of Frequentist and Bayesian Methods on
Survey Sampling Practice: A Selective Appraisal by J. N. K. Rao}

\begin{aug}
\author[a]{\fnms{J.} \snm{Sedransk}\corref{}\ead[label=e1]{jxs123@cwru.edu}}
\runauthor{J. Sedransk}

\affiliation{Case Western Reserve University}

\address[a]{J. Sedransk is Professor (Emeritus), Department of Statistics, Case Western Reserve
University, 10600 Crossing Creek Road, Potomac, Maryland 20854, USA \printead{e1}.}

\end{aug}

\vspace*{-12pt}
\begin{abstract}
This comment emphasizes the importance of model checking and
model fitting when making inferences about finite population quantities.
It also suggests the value of using unit level models when making
inferences for small subpopulations, that is, ``small area'' analyses.
\end{abstract}

\begin{keyword}
\kwd{Diagnostics}
\kwd{hierarchical structure}
\kwd{model checking}
\kwd{model fitting}
\kwd{small area statistics}
\kwd{unit level models}.
\end{keyword}

\end{frontmatter}

Professor Rao has written an excellent review of the alternative methods
of making inference for finite population quantities. This is an
underserved field of research and, hopefully, this paper will encourage
some readers to make contributions to this important, practical area.

Rather than commenting on detailed aspects of the paper, I will discuss
two broad areas. Both are treated briefly in this article, but have not
been considered in the survey sampling literature as fully as I~think
they should be. The first is the fitting of models to complex survey
data, and the second is model checking.

Except for the design-based approach, all of the inferential methods
described in this paper rely significantly on models. And, over the past
thirty years great strides have been made to develop models that are
consistent with observed data. My impression, though, is that survey
statisticians have been slow to adopt these methodological advances. In
Section~1 Rao writes, referring to Hansen, Madow and\vadjust{\eject} Tepping (\citeyear{HANMADTEP83}),
``Unfortunately, for large samples [mo\-del dependent approaches] may
perform very poorly under model misspecifications; even small model
deviations can cause serious problems.'' This example (in Hansen, Madow and Tepping, \citeyear{HANMADTEP83}) was analyzed almost thirty years ago and \textit{only by the
authors}. One would hope that current methodology and skills in data
analysis would provide an improvement over the Hansen, Madow and Tepping (\citeyear{HANMADTEP83})
``straw man,'' the usual ratio estimator. As noted by Hansen, Madow and Tepping (\citeyear{HANMADTEP83}), one should use robust methods. But, there have been other
advances in diagnostic techniques and inferential methods (e.g., model
averaging). Moreover, this is a \textit{single} example and, before
drawing general conclusions, it would be preferable to consider this
example again and analyze other examples typical of sample survey data.
Finally, though, it is important to note that there are challenging
problems in modeling data from complex sample surveys because there may
be several stages of cluster sampling, small sample sizes (typically in
inconvenient places), possible selection biases, nonresponse and
measurement errors.

When the objective is inference for ``small area'' quantities there are
special issues with modeling. In my experience almost all of the
\textit{applications} use an area-level model; see, for example,
Section 5 of this paper and Rao (\citeyear{Rao03}). (Moreover, there are many
applications that are not reported in the refereed literature, and I do
not know of any that use\vadjust{\eject} a~unit-level model.) In a small area analysis
one is concerned about the quality of the direct estimator, $\hat{\theta}_i$, and,
thus, uses a model that adds information about other small areas to
improve inference about $\theta_i$. Clearly, then, the quality of the
estimated variance of $\hat{\theta}_i$, $v(\hat{\theta}_i)$, is even more questionable. (Rao
notes this in Section 5, i.e., ``the second assumption of known sampling
variances is more problematic.'') Moreover, is it reasonable to assume
that $(\hat{\theta}_i-\theta_i)/\sqrt{v(\hat{\theta}_i)}$ is satisfactorily approximated by a standard normal
distribution? A transformation of $\hat{\theta}_i$ may be helpful. But, choosing
the transformation and verifying that the associated standardized
quantity is approximately distributed as $N(0,1)$ is a challenging
exercise. There is a better way, though, and that is to model the
\textit{unit} level data as, for example, in Battese, Harter and Fuller
(\citeyear{BATHARFUL88}), Malec, Sedransk, Moriarity and LeClere (\citeyear{MALetal97}) and Malec (\citeyear{MAL05}).
Doing so has a second benefit. In such circumstances one can investigate
alternative ways to make inference about the $\theta_i$ from an \textit{area-level} model (because the microdata are now available and one can
investigate sampling distributions of the transformed
$\hat{\theta}_i$'s).

Model checking is an essential part of the mode\-ling process. In Section
5, Rao writes that ``some~of the default HB model-checking measures that
are~wi\-dely used may not be necessarily good for \mbox{detecting} model
deviations. For example, the commonly used posterior predictive
$p$-value (PPP) for checking good\-ness-of-fit may not be powerful enough
to detect non-normality of random effects\ldots\ because this measure makes
`double use' of the data\ldots.'' There are methods that take care of this
problem, for example, the partial PPP and conditional PPP (Bayarri and Berger, \citeyear{BayBer00}), and the newer CPPP (Hjort, Dahl and Steinbakk, \citeyear{HjoDahSte06}).
While these are computationally intensive, this should not be a major
limitation in the current era. (See Ma, Sun and Sedransk, \citeyear{MASUNSED}, for a
recent implementation of CPPP.) I think, though, that there are other
considerations that are probably even more important. First, choosing
the appropriate test quantities to assess the fit of the currently
entertained model is essential. And, this is difficult because an
appropriate selection depends on guessing the nature of the aberration
of the currently entertained model from one that is closer to the one
that generated the observed data. See, for example, Yan and
Sedransk (\citeyear{YANSED06}, \citeyear{YanSed07}, \citeyear{YANSED10})
who
investigated in detail the problem of detecting
unknown hierarchical structure (e.g., fitting a model\vadjust{\eject} with a single
stage when, in actuality, there are two stages). Moreover, is it
important to detect relatively small discrepancies from the model
currently being entertained? One may be requiring more ``po\-wer'' than is
warranted by the intended use of the data. Additionally, tests of
goodness-of-fit are problematic, especially in the frequentist paradigm
since such tests are constructed to \textit{reject} null hypotheses
whereas one would like to accept a postulated model if the data are
concordant with it.

Finally, in Sections 4 and 5, Rao has discussed some applications of
Bayesian methods to sample survey data. Sedransk (\citeyear{SED08}), referenced in
Rao's paper, describes other areas where the use of Baye\-sian techniques
should be useful, and also points out some limitations.


\end{document}